\documentstyle[12pt,aaspp]{article}

\newcommand{\cobedmr}{{\em COBE}/DMR\ }
\newcommand{\etal}{et~al.\ }
\newcommand{\microk}{~\mu{\rm K} }
\newcommand{\maxn}{1.0}
\newcommand{\maxQ}{19~\microk }
\newcommand{\nsims}{20000}
\newcommand{\abshin}{2.2}
\newcommand{\abslon}{-0.4}
\newcommand{\abshiQ}{44~\microk }
\newcommand{\absloQ}{8~\microk }
\newcommand{\connhiQ}{25~\microk }
\newcommand{\connloQ}{14~\microk }
\newcommand{\conQhin}{1.4}
\newcommand{\conQlon}{0.5}
\newcommand{\favoriteR}{430~(\microk)^2}
\newcommand{\favoriteNmax}{171}

\begin{document}

\title{The Amplitude and Spectral Index of the Large Angular
       Scale Anisotropy in the Cosmic Microwave Background
       Radiation}

\author{
   KEN GANGA\altaffilmark{1}, LYMAN PAGE\altaffilmark{2}
      \vspace{8pt}
      \affil{Joseph Henry Laboratories,
             Physics Department,
             Jadwin Hall, Princeton University,
             Princeton, NJ \ 08544--0708}
   EDWARD CHENG\altaffilmark{3}
      \vspace{8pt}
      \affil{NASA/Goddard Space Flight Center, Code 685.0,
             Greenbelt, MD \ 20771}
   STEPHAN MEYER\altaffilmark{4}
      \vspace{8pt}
      \affil{Room Acc F100, Enrico Fermi Institute,
             University of Chicago, 5640 South Ellis Ave.,
             Chicago, IL \ 60637}
}
\altaffiltext{1}{E-Mail: ganga@pupgg.princeton.edu}
\altaffiltext{2}{E-Mail: page@pupgg.princeton.edu}
\altaffiltext{3}{E-Mail: ec@stars.gsfc.nasa.gov}
\altaffiltext{4}{E-Mail: meyer@oddjob.uchicago.edu}

\begin{abstract}

In many cosmological models, the large angular scale anisotropy
in the cosmic microwave background is parameterized by a spectral
index, $n$, and a quadrupolar amplitude, $Q$. For a
Peebles-Harrison-Zel'dovich spectrum, $n=1$. Using data from the
Far Infra-Red Survey (FIRS) and a new statistical measure, a
contour plot of the likelihood for cosmological models for which
$-1 < n < 3$ and $0 \le Q \le 50~\mu{\rm K}$ is obtained. We find
that the likelihood is maximum at $(n, Q) = (\maxn,\maxQ)$. For
constant $n$ the likelihood falls to half its maximum at $Q
\approx \connloQ$ and $\connhiQ$ and for constant $Q$ the
likelihood falls to half its maximum at $n \approx \conQlon$ and
$\conQhin$. Regardless of $Q$, the likelihood is always less than
half its maximum for $n < \abslon$ and for $n > \abshin$, as it
is for $Q < \absloQ$ and $Q > \abshiQ$.

\end{abstract}


\section{Introduction}

With the \cobedmr detection of anisotropy in the cosmic microwave
background radiation (CMB)~(\cite{Bennett92}; \cite{Smoot92};
\cite{Wright92}), followed by the FIRS--\cobedmr
cross-correlation~(\cite{Ganga93}) and a host of recently
reported detections on smaller angular scales, we must now
characterize the nature of the detected variations.

The inflationary scenario predicts a spectrum of primordial
density perturbations of form $P(k) \propto k$ for large angular
scales (\cite{Steinhardt84}), while popular variants of cold dark
matter (CDM) models predict $P(k) \propto k^n$ with $0.7 < n < 1$
(see Turner (1993) for a review). Textures will also appear
scale-invariant (i.e., $n = 1$; \cite{Pen93}), while cosmic
strings may mimic an $n$ of
1.35--1.5~(\cite{Perivolaropoulos94}). Thus, by limiting the
range of $n$ allowed by the data we can constrain theories of
structure formation. It should be pointed out, however, that many
theories are not well parameterized this way~({\em e.g.}, Ratra
\& Peebles (1994)) and that in this work we consider models with
{\em Gaussian} fluctuations only.

If one assumes this power law spectrum for density perturbations
in an Einstein-DeSitter universe, then the angular power spectrum
of anisotropies in the CMB is given by (\cite{Bond87})
\begin{equation}
C_l = Q^2{4\pi\over 5}{\Gamma\Bigl((2l+n-1)/2\Bigr)
       \Gamma\Bigl((9-n)/2)\Bigr)\over
       \Gamma\Bigl((2l+5-n)/2\Bigr)
       \Gamma\Bigl((3+n)/2)\Bigr)}, \label{eq:csubl}
\end{equation}
where $Q = \sqrt{5C_2/4\pi}$ is a parameter that sets the
amplitude of the theoretical angular power spectrum and
correlation function at the angular scale of the
quadrupole.\footnote{Smoot et~al. (1992) use the notation
$Q_{rms-PS}$ to denote the parameter, but for the sake of brevity
we follow the notation of Wright et~al. (1994b) and use $Q$. Note
that $Q$ is {\em not} the root-mean-squared amplitude of the
quadrupole in our Universe; we do not measure the intrinsic
quadrupole because of our limited sky coverage.} Note also that
Equation~\ref{eq:csubl} does not account for microphysical
processes that can change the power spectrum of anisotropies and
thus alter the effective $n$; for example, ``standard'' scale
invariant CDM has an effective $n$ of 1.15 for  $l {_<\atop^\sim}
30$~(\cite{Bond94b}).

The first estimate of $n$ using large scale CMB anisotropy data
was made by fitting the theoretical auto-correlation function
(ACF) derived from Equation~\ref{eq:csubl} directly to the
auto-correlation function of the \cobedmr first year data. The
result was $n = 1.15_{-0.65}^{+0.45}$ and $Q = 16.3\pm
4.6~\mu{\rm K}$~(\cite{Smoot92}). Seljak and Bertschinger (1993),
using a maximum-likelihood analysis again on the first-year
\cobedmr data, found that $Q = (15.7\pm 2.6)e^{0.46(1-n)}~\mu{\rm
K}$. Additionally, Smoot et~al. (1994) and Torres (1994) have
both done topological analyses of these data to limit $n$ to
$1.7_{-0.6}^{+0.3}$ and $1.2 \pm 0.3$, respectively. Bond (1994)
has extracted the index from an analysis of the power spectrum of
both FIRS and DMR. He finds $n = 1.8_{-0.8}^{+0.6}$ for FIRS and
$n = 2.0_{-0.4}^{+0.4}$ for the first-year DMR maps. Wright
et~al. (1994b), using a power spectrum analysis with the two-year
DMR maps, find $n = 1.46_{-0.44}^{+0.41}$. Bennett et~al. (1994),
also using the DMR two-year maps, perform a maximum-likelihood
analysis with Monte Carlo simulations and find $n =
1.59_{-0.55}^{+0.49}$. Finally, G\'orski et~al. (1994) use a
power spectrum based method to compute the likelihood of models.
Applying it to the DMR two-year data, they obtain a maximum
likelihood at $(n, Q) = (1.2, 17 \microk)$ and a maximum in the
marginal likelihood for $n$ at $1.1 \pm 0.3$.

This range of results, often with the same data, indicates the
difficulties involved in the analyses. There is clearly a
need for more data and consistent statistical techniques. In this
{\it Letter}, we describe a method for analyzing CMB anisotropy
maps and apply it to the Far Infra-Red Survey (FIRS). This method
uses the correlation function of anisotropy maps. Thus, it is
relatively insensitive to noise contamination that can infect
analyses based upon the $\chi^2$ or the power spectrum of a
map~(\cite{Bond94b}).

\section{The Far Infra-Red Survey}

FIRS is a balloon-borne, bolometric anisotropy experiment that
observes in four channels at 170, 290, 500  and 680 GHz. It has a
beam full-width-at-half-maximum (FWHM) of $3.\deg 8$ and has
been flown successfully twice, resulting in coverage over most of
the Northern hemisphere. The data used here are from the 170 GHz
channel of the October, 1989 flight and cover roughly one quarter
of the sky. The data are shown in Ganga et al. (1993). The
experiment and map are described more fully in Meyer et~al.
(1991) and Page et~al. (1989, 1990, 1992, 1994a,b).

For the purposes of this {\em Letter}, it is sufficient to note
that the map is a set of pixels $i$, each with an associated
residual temperature $t_i$ and statistical weight  $w_i$. The
$t_i$'s are the average temperatures of all measurements made
within pixel $i$ after the best fit offset, dipole and Galactic
model (the {\sl IRAS} $100~\mu$m map smeared to a 3.\deg 8 beam
width) are removed. The statistical weights are the reciprocals
of the variances of the measurements for each pixel. A 15\deg\
Galactic latitude cut has been imposed on these data and on the
simulations described below to minimize the effects of residual
dust emission near the Galactic plane. For this analysis, we
assume that all the correlations in the map are due to
correlations in the CMB. In Page et al. (1994b), the data
reduction and known systematic errors are discussed. There are no
{\em known} systematic effects that introduce correlations into
the data at levels which affect	the results of this analysis.

\section{Analysis}

The correlation function of the CMB is estimated from the data
with
\begin{equation}
   C(\theta_A) = {1\over\Gamma_A}\sum_{i,j\in A} w_iw_jt_it_j,
\end{equation}
where the sum is over all pixels $i$ and $j$ for which the angle
between them is $\theta_A$ and where
\begin{equation}
   \Gamma_A = \sum_{i,j\in A} w_iw_j.
\end{equation}
$C(\theta_A)$ is presented in Figure 1. The correlation function
is divided into 64 2.\deg 8 bins, of which, because these data do
not cover the entire celestial sphere, only $n_c = 58$ contain
data. Also, the correlation function at $\theta=0$ (or,
alternately, the {\em rms} of the data) is excluded from the
analysis, as it is affected most by instrumental noise.

Simulations of the sky are made for each $n$ from 0.9 to 2.9 in
steps of 0.1 following the methods of Cottingham (1987) and
Boughn et~al. (1993). Specifically, maps of unit amplitude are
made with underlying temperatures at each pixel $i$ defined by
\begin{equation}
   T_i = \sum_{l=2}^{l_{max}}\sum_{m=-l}^l
      a_{l,m}\sqrt{W_l}Y_{l,m}(\theta_i,\phi_i),
\end{equation}
where $W_l$ is related to the Legendre transform of the
experiment's beam response. If one assumes that the response is
Gaussian, as we do here (but see Wright et al. (1994a)), this
window function can be approximated by $W_l = e^{-0.18\cdot
l(l+1)\theta_{1/2}^2}$, where $\theta_{1/2}$ is the FWHM of the
beam (in radians). With a FWHM of 3.\deg 8, $W_l$ falls to 0.5 at
$l = 29$. For these simulations, $l_{max}$ is either 150 or the
highest $l$ for which $W_lC_l/W_2C_2 > 10^{-7}$, whichever is
lowest. Each $a_{l,0}$ is drawn from a normal distribution with a
variance of $C_l/(0.8\pi Q^2)$ and for $m > 0$ both the real and
imaginary parts of $a_{l,m}$ are drawn from normal distributions
with variances of $C_l/(1.6\pi Q^2)$. Finally, $a_{l,-m} = (-1)^m
a_{l,m}^*$. We call these signal maps. In order to include
experimental uncertainties, separate noise maps are created. The
temperature at each pixel $i$ is drawn from a normal distribution
with a variance of $1/w_i$. The pixels in both the signal and
noise maps have the same weights as the real data. In order to
match the processing done on the real data, an offset and a
dipole are removed from each map separately in a least-squares
fit.

Note that if the maps contain nothing but statistical noise,
$\Gamma_A$ is the reciprocal of the variance in bin $A$. That is,
$\sigma^2_{C(\theta_A)} = {1/\Gamma_A}$~(see Ganga et~al. (1993)
for a derivation and Smoot et~al. (1994) for a discussion of
weighting schemes for different statistics over CMB maps).

With this prescription, we form $C_{SS}(\theta)$, the ACF of a
signal map, $C_{SN}(\theta)$, the cross-correlation between a
signal map and its associated noise map, and $C_{NN}(\theta)$,
the ACF of a noise map. We note that the full ACF of a single
sky realization with a particular value of $Q$ and noise will be
\begin{equation}
   C(\theta) = Q^2{4\pi\over 5}C_{SS}(\theta)
      + 2Q\sqrt{4\pi\over 5}C_{SN}(\theta) + C_{NN}(\theta).
\end{equation}
This scaling allows one to reduce the number of simulations made,
though a set of simulations at each value of $n$ to be tested
is still required.

For each simulation we make the statistic
\begin{equation}
   R = \sqrt{{1\over \sum_A \Gamma_A}
       \sum_A \Gamma_A \Bigl(C_F(\theta_A) -
                                   C_D(\theta_A)\Bigl)^2},
\end{equation}
where $C_F(\theta_A)$ is the value of the ACF of a simulation at
bin $A$ and $C_D(\theta_A)$ is the value of the ACF of the data
at correlation bin $A$. If $R = 0$, the ACF of the simulation is
the same as that of the data.

We calculate the likelihood, $P(D|n,Q)$, or the relative
probability of obtaining the FIRS data ACF given a model
parametrized by $n$ and $Q$, by choosing a limiting value,
$R_{lim}$, for $R$ and finding the number of simulations, $N$,
for which $R < R_{lim}$. We normalize the distribution so that
the maximum likelihood is unity. The {\em a posteriori}
probability that the Universe is parameterized by a certain $n$
and $Q$ can then be found by invoking Bayes's theorem with a
suitable {\em a priori} probability distribution for $(n, Q)$
(see, for example, Martin (1971)).

In essence, we are treating the correlation functions as vectors
in an $n_c$ dimensional ACF space. The various $R$'s represent
the weighted magnitude of the difference between the data ACF
vector and a simulation ACF vector. If the $\Gamma_A$'s were all
the same, $R_{lim}$ would define the radius of an $n_c$
dimensional sphere centered upon the data ACF. When the
$\Gamma_A$ are different, the sphere is deformed into an
ellipsoid. By counting the number of simulations with $R$ less
than some chosen $R_{lim}$, we are calculating the density of
simulated ACF's in the neighborhood of the measured ACF. This
density is proportional to the likelihood. The method accounts
for non-uniform sky coverage, instrumental noise, cosmic variance
and the effects of removing an offset and dipole from an the
data.

\section{Results and Discussion}

Figure~2 shows the results of the analysis for $-1 < n < 3$ and
$0~\mu{\rm K} < Q < 50~\mu{\rm K}$. In this case, $R_{lim} =
\favoriteR$, resulting in $N_{max} = \favoriteNmax$ simulations
at the most likely combination of $(n, Q)$ with $R \le R_{lim}$.
That is, for $(n,Q) = (1.0, 19~\microk)$, 171 out of a total of
20000 simulations had $R$ less than $\favoriteR$. All other
combinations of $(n,Q)$ had fewer simulations fall within the
test volume. Figure~2 shows likelihood contours for 0.05, 0.25,
0.5 and 0.75 with solid lines along with a large $\times$ at the
maximum. The inner broken line marks the $P(D|n,Q) = e^{-1/2}$
likelihood contour, while the outer broken line corresponds to
$P(D|n,Q) = 0.34$. They are shown for comparison to other
analyses. If one were to extract only the $n = 1$ portion of
Figure~2, in the limit that the extracted likelihood is Gaussian,
the $P(D|n,Q) = e^{-1/2}$ points would equal the $1\sigma$
credible limits assuming a uniform prior. The $P(D|n,Q) = 0.34$
contour represents the 68\% credible limits if we use a prior of
one for the models considered here and zero for all others. In
other words, the the sum of the likelihoods within the 0.34
contour is 68\% of the sum of all the likelihoods in Figure~2.

For each model, the number of simulations within our limiting
radius is governed by the binomial distribution. The probability
of a particular simulation falling within the test volume is much
smaller than that of it falling outside the volume. We can,
therefore, approximate the variance in the estimate of the
various $N$'s by $N$ itself. Hence, the uncertainty in the
probability for those models with $P(Q,n|D) \approx 1$ is
approximately $\sqrt{2/N_{max}}$, and the error in the estimates
of the likelihoods near the maximum likelihood in Figure 2 is
approximately 11\% (with $N_{\rm max} = 171$). This is borne out
by Figure 3, which shows that the $(n,Q)$ with the maximum
likelihood depends slightly on the values of $R_{lim}$, but not
by amounts in excess of what is expected statistically.

Bond (1994) has noted that the FIRS data contain ``white noise.''
In principle, this noise affects only the  $\theta_A = 0$ bin of
the correlation function, which is not included in this analysis.
To check the possibility that the noise has `leaked' out of the
zero ACF bin, the analysis was repeated after excising the bin at
$\theta_A = 1.\deg 4$ (effectively eliminating all correlations
on angular scales less than 2.\deg 8). The results are consistent
with those quoted above. If the data contained only white noise,
Figure~2 would have a maximum at n = 3. Clearly, the white noise
does not significantly contaminate these results. Again, this is
because this method is primarily sensitive to spatial
correlations in the data.

This method has been checked in a number of ways. Figure 3 shows
that there are small changes in the likelihood contours as
$R_{lim}$ is increased. This indicates that a smaller value of
$R_{lim}$ is desirable. With this number of simulations, however,
reducing $R_{lim}$ degrades the plot to the point where it is no
longer useful. Note though that the likelihood is stable for $n
\approx 1$ and that for other $n$'s the likelihood values are
conservative (that is, the likelihood is {\em overestimated}).

We have also checked that changing the weighting in the
definition of $R$ does not affect the results. Redefining $R$
such that
\begin{equation}
   R = \sqrt{{1\over \sum_A W_A}
         \sum_A W_A \Bigl(C_S(\theta_A) -
                                   C_D(\theta_A)\Bigl)^2},
\end{equation}
where $W_A$ is now defined as 1, $\sqrt{\Gamma_A}$, $\Gamma_A^2$
or $\Gamma_A^3$ yields consistent results, though, the
bounds are not as strict.

We put limits on the bias of this method by applying the
procedure to 500 additional simulations with $(n, Q) = (1.0,
20~\microk)$. Though not a well defined quantity because some of
the simulations would naturally prefer an $n < -0.9$ or an $n >
2.9$, the mean result, $(n, Q) = (0.8, 16.5~\microk)$, is
consistent with the input values to within the uncertainties in
estimating the mean.

The approximation of the FIRS beam by a $3.\deg 8$ Gaussian is
adequate. Results from simulations made with a Legendre transform
of the measured profile do not differ substantially. If a
Gaussian of width $4.\deg 2$ is assumed to account for smearing,
then the value for $Q$ at $n=1$ increases by $1~\mu$K.

Figure 2 implies that for the FIRS data $n \approx 1$ is favored.
We point out, however, that only a fraction of the total ACF
space has been tested. Many other possibilities exist. Wright
\etal\ (1992) found the ACF of the \cobedmr first-year data to be
well described by a Gaussian with a coherence angle, $\theta_c$,
of $13.\deg 5$; that is, the correlation function of the DMR data
was fit well by a function of form $C(\theta) =
C(0)e^{-0.5(\theta /\theta_c)^2}$ with adjustments for offset and
dipole removal, where $C(0)$ is the variance of the intrinsic sky
fluctuation. This statistical method is well suited to testing
this possibility. Retaining the normalization used in Figure 2,
we find the likelihood of such a model with $C(0) \approx
(44~\microk)^2$ to be 1.4. That is, 40\% more likely than the
most likely model based on a power law spectrum. Thus, while $n
\approx 1$ may be preferred compared to other models with a power
spectrum of density perturbations that follows a power law, we
cannot prove it is the best model.

We emphasize that no one experiment can prove that the correlated
signal measured is due solely to the CMB. However, in the case of
the FIRS data we have ruled out correlation stemming from
astrophysical foregrounds, the instrument, and analysis artifacts
(Page~et~al. 1994b). The similarity of this result to the COBE/DMR
auto-correlation function and the significant cross-correlation
(Ganga~et~al. 1993) with DMR strongly suggest that both
experiments are detecting correlations in the CMB. Work is
underway to generalize our method to find the most likely (n,Q)
for the FIRS/DMR cross-correlation.

\section{Acknowledgements}

We would like to thank Steve Boughn and Dick Bond for insights
and suggestions. We also thank E.~Bertschinger, D. Cottingham,
K.~G\'orski,  G.~Hinshaw, B.~Ratra, and N.~Turok for helpful
comments and conversations.  KMG would also like to thank Ed Hsu
for insights into Monte Carlo methods. This work was supported
under NASA grants NAGW--1841 and NAG5--2412, NSF grant PH
89--21378, and an NSF NYI grant to L. Page.

\newpage

\section{Figure Captions}

\begin{figure}[h]
   \caption{The FIRS auto-correlation function. The error bars
         represent the variations due to experimental noise only.
         For this plot, the $180\deg$ span of $\theta_A$ is
         divided into 40 bins each of width $4.\deg 5$. Due to
         the limited sky coverage, only 36 bins contain data. The
         shaded area shows the 1$\sigma$ spread in the \nsims\
         simulated auto-correlation functions for $n = 1$ with an
         amplitude of $Q = 19~\mu K$. The broken curve
         (corresponding to the right axis) shows the number of
         pairs of pixels contributing to the correlation function
         at each correlation function bin.}
\end{figure}

\begin{figure}[h]
   \caption{Likelihood contours for $R_{lim} = 430~(\microk)^2$,
         corresponding to a maximum of 171 simulations included
         within the test volume. The solid contours correspond to
         likelihoods of 0.05, 0.25, 0.5, and 0.75. The $\times$
         at $(n,Q) \approx (1.0,19~\microk )$ is the maximum and
         the broken lines correspond to likelihoods of 0.34 and
         0.68.}
\end{figure}

\begin{figure}[h]
   \caption{Likelihood contours for various values of $R_{lim}$.
         The values of $R_{lim}$ are 410, 420, 430 and
         440~$(\microk)^2$, moving clockwise from the top left.
         The maximum number of simulations within the test
         volumes for each case is indicated on the plots.}
\end{figure}


\begin{thebibliography}

\bibitem[Bennett et~al. 1992]{Bennett92}
Bennett, C.L. et~al. 1992,
\newblock \apj, 396(1):L7--L12

\bibitem[Bennett et~al. 1994]{Bennett94}
Bennett, C.L. et~al. 1994,
\newblock submitted to \apj

\bibitem[Bond 1994]{Bond94b}
Bond, J.R. 1994,
\newblock In {\em Proceedings of the 1994 Capri CMBR Conference}

\bibitem[Bond \& Efstathiou 1987]{Bond87}
Bond, J.R. \& Efstathiou, G. 1987,
\newblock \mnras, 226:665

\bibitem[Boughn et~al. 1993]{Boughn93}
Boughn, S.P. et~al. 1993,
\newblock \apj, 391(2):L49--L52

\bibitem[Cottingham 1987]{Cottinghamthesis}
Cottingham, D. 1987,
\newblock PhD thesis, Princeton University

\bibitem[Ganga et~al. 1993]{Ganga93}
Ganga, K. et al. 1993,
\newblock \apj, 410:L57

\bibitem[G\'orski et al. 1994]{Gorski94}
{G\'orski}, K. et~al. 1994,
\newblock submitted to ApJ

\bibitem[Martin 1971]{Martin}
Martin, B.R. 1971,
\newblock {\em Statistics for Physicists}, Academic Press

\bibitem[Meyer et~al. 1991]{Meyer91}
Meyer, S.S., Cheng, E.S. \& Page, L.A. 1991,
\newblock \apj, 371(1):L7--L9

\bibitem[Page 1989]{Pagethesis}
Page, L.A. 1989,
\newblock PhD thesis, Massachusetts Institute of Technology

\bibitem[Page et~al. 1994a]{Page94a}
Page, L.A. et~al. 1994a,
\newblock {\em Applied Optics}, 33(1):11--23

\bibitem[Page et~al. 1994b]{Page94b}
Page, L.A. et~al. 1994b,
\newblock In {\em Astrophysical Letters and Communications:
   Proceedings from the 1994 Capri CMBR Workshop}

\bibitem[Page et~al 1990]{Page90}
Page, L.A., Cheng, E.S. \& Meyer, S.S. 1990
\newblock \apj, 355(1):L1--L4

\bibitem[Page et~al. 1992]{Page92}
Page, L.A., Cheng, E.S. \& Meyer, S.S. 1992
\newblock {\em Applied Optics}, 31(1):95--100

\bibitem[Pen et~al. 1993]{Pen93}
Pen, U., Spergel, D. \& Turok, N. 1994,
\newblock {\em Phys. Rev. D}, 49, 2

\bibitem[Perivolaropoulos 1993]{Perivolaropoulos94}
Perivolaropoulos, L. 1993,
\newblock Center for Astrophysics preprint CfA--3796

\bibitem[Ratra \& Peebles 1994]{Ratra94}
Ratra, B. \& Peebles P.J.E. 1994,
\newblock Princeton University Physics preprint PUPT-1444;
          Submitted to {\em Phys. Rev. D}

\bibitem[Seljak and Bertschinger 1993]{Seljak93}
Seljak, U. \& Bertschinger, E. 1993,
\newblock \apj, 417(1):L9--L12

\bibitem[Smoot et~al. 1992]{Smoot92}
Smoot, G.F. et~al 1992,
\newblock \apj, 396(1):L1--L5

\bibitem[Smoot et~al. 1994]{Smoot94}
Smoot G.F. et~al. 1994,
\newblock submitted to \apj

\bibitem[Steinhardt \& Turner 1984]{Steinhardt84}
Steinhardt P.J. \& Turner, M.S. 1984,
\newblock {\em Physical Review D}, 29:2162

\bibitem[Torres 1994]{Torres94}
Torres, S. 1994,
\newblock \apj, 423(1):L9--L12

\bibitem[Turner 1993]{Turner93}
Turner, M.S. 1993,
\newblock FERMILAB-Conf-92/313-A

\bibitem[Wright et~al. 1992]{Wright92}
Wright, E.L. et~al. 1992,
\newblock \apj, 396(1):L13--L18

\bibitem[Wright et~al. 1994a]{Wright94a}
Wright, E.L. et~al. 1994a,
\newblock \apj, in press

\bibitem[Wright et~al. 1994b]{Wright94b}
Wright, E.L. et~al. 1994b,
\newblock submitted to \apj

\end{thebibliography}
\end{document}